\documentclass[12pt,american,english]{article}
\usepackage[latin9]{inputenc}
\usepackage{amsmath}
\usepackage{amssymb}
\usepackage{graphicx}
\usepackage{esint}

\makeatletter

\providecommand{\tabularnewline}{\\}

\newcommand{\lyxaddress}[1]{
\par {\raggedright #1
\vspace{1.4em}
\noindent\par}
}

\@ifundefined{definecolor}{\usepackage{color}}{}

\usepackage{babel}

\makeatother

\usepackage{babel}
\begin{document}

\title{Description of the plasma delay effect in silicon detectors}

\author{Z. Sosin}

\maketitle

\lyxaddress{Institute of Physics, Jagellonian University}
\begin{abstract}
A new method of modeling of the current signal induced by charged
particle in silicon detectors is presented. The approach is based
on the Ramo-Shockley theorem for which the charge carrier velocities
are determined by taking into account not only the external electric
field generated by the electrodes, but also the Coulomb interaction
between the electron and hole clouds as well as their diffusion. 
\end{abstract}
\emph{keywords: }

Plasma delay

Current pulse

Pulse shape analysis

Particles identification

Silicon detector

\section{Introduction}

It is obvious that identification of particles and fragments produced
in nuclear reactions is crucial for any experimental or technical
work in nuclear physics. Among different ways of identifying charged
particles the classical $\Delta E-E$ telescope method remains still
the flagship. Recently, an alternative method based on the Pulse Shape
Discrimination (PSD) technique applied for silicon detectors is being
developed and is increasingly drawing attention. Recent results demonstrate
that the method can offer charge and isotopic identification comparable
to that obtained with the classical $\Delta E-E$ method. The main
advantage of the PSD method comes from the fact that it requires only
one electronic channel for detection and identification. It is thus
an important point for designing and constructing multi-detector systems.

A significant difference between the $\Delta E-E$ and the PSD techniques
results from the fact that the former is governed basically by the
energy loss process (Bragg curve), while the latter is primarily related
to the Plasma Delay Effect (PDE) {[}1-6{]}. In silicon detectors,
this effect manifests itself with shortening of the pulse rise time
with decreasing $Z$ for low and intermediate mass fragments, for
which the generated charge is practically completely collected by
the detector electrodes. The experimental data demonstrates that the
PDE concerns particles with small $Z$, for which the Pulse Hight
Defect (PHD) is still of little importance.

For better understanding of the identification idea associated with
PSD technique, and for its future development, it is crucial to have
at ones command a perfect simulation of the time dependence of the
experimental signal produced by an ion with a given charge, $Z$,
atomic mass, $A$, and energy, $E$. The main goal of such a simulation
is to describe the extraction and collection of the generated charge
carriers moving in the external electric field distorted due to the
presence of a highly ionized track and due to the diffusion process
of the carriers.

As presented in \cite{Harmita_04}, an approach in which the distortion
of the electric field caused by the generated carriers is neglected,
is able to correctly describe the current signals for light charged
particles (LCP), e.g. protons. However, this simplified approach completely
fails in case of heavy ions (HI) for which the collection time of
the generated carriers gets longer $(\tau^{HI}>\tau^{LCP})$. Historically,
this difference, associated to a slower carrier collection for HI,
was quantified as a plasma delay (PD) effect. Since this effect influences
the current signal rise-time, it appears to be crucial for the PSD
technique.

An attempt to describe phenomenologically the delayed carrier collection
time in silicon detectors has been recently proposed in \cite{Parlog_10}.
The proposed description took into account the polarization of the
electron-hole pairs generated by the HI and connected it to the relative
dielectric permittivity. Another important assumption was that the
dissociation of pairs in time occurred with a constant probability
and the modified electric field, inside and outside of the ion range,
was given by the Maxwell equation for the electric field in the inhomogeneous
medium. With these assumptions, the model was indeed able to describe
the experimental pulse shapes quite accurately.

In the present paper we propose another, more microscopic approach.
The main model assumptions are the following: 
\begin{quotation}
i. Propagation of the electric charges (electrons, holes) generated
in the detector is represented by evolution of the Gaussian clouds
for which the centroids and variances are treated as independent variables.

ii. Position of the centroid of each Gaussian is governed by the drift
process, while its variance undergoes both, diffusion and the drift
process. 
\end{quotation}
The first results of the model calculations indicate some binding
effects between the holes and electrons, in a region similar to that
predicted by the phenomenological model of Ref. \cite{Parlog_10}.
In this region also the electric field shows similar behavior to that
presented in \cite{Parlog_10}.

The detailed description of the new model is presented in the following
section. The preliminary results of the calculations and comparison
to the experimental data are presented in Section 3; Conclusions and
possible extensions of the model applicability are given in Section
4.

\section{Description of the model}

A particle entering the silicon detector is assumed to degrade its
energy according to the Bragg curve which relates the generated ionization
$B(x)$ to the particle position $x$. We assume that the X direction
is perpendicular to the detector surface. In order to describe the
initial, local, density of the electrons $\rho_{e}(\mathbf{r,}t=0)$
and holes $\rho_{h}(\mathbf{r,}t=0)$ we assume that the ionization
is proportional to the local stopping power $B(x)=\frac{1}{w}\frac{dE}{dx}(x)$,
where $w=3.62$ eV is the energy for an electron-hole pair production
and $\frac{dE}{dx}(x)$ is the local stopping power \cite{Zigler_85}.
Just after stopping of the impinging ion, the carrier density can
be described as:

\begin{equation}
\rho_{e}(\mathbf{r,}t=0)=-\rho_{h}(\mathbf{r,}t=0)=-\int B(x')\delta(x-x')\delta(y)\delta(z)\, dx'\label{eq: jonizacja}
\end{equation}
 where $\mathbf{r}=\left[x,y,z\right]$. This assumption states that
for $t=0$ the ionization is localized along the X axis only and disappears
elsewhere.

In order to describe the time evolution of the generated ionization
we assume its distribution in the following form:

\begin{equation}
\rho_{e}(\mathbf{r,}t)=\int B_{e}(x',t)G_{e}(x-x',y,z,t)\, dx'\label{eq:ro_e}
\end{equation}
 
\begin{equation}
\rho_{h}(\mathbf{r,}t)=\int B_{h}(x',t)G_{h}(x-x',y,z,t)\, dx'\label{eq:ro_h}
\end{equation}
 which is analog to (\ref{eq: jonizacja}) and we set: 
\begin{equation}
-B_{e}(x,t=0)=B_{h}(x,t=0)=B(x)\label{eq:Bt0}
\end{equation}

Functions $G_{e}$ and $G_{h}$ are assumed to be Gaussians: 
\begin{equation}
G_{e}(x-x',y,z,t)=\frac{1}{\sqrt{\left(2\pi\right)^{3}}\sigma_{e}^{3}\left(x,t\right)}\exp\left(-\frac{\left(x-x'\right)^{2}+y^{2}+z^{2}}{2\sigma_{e}^{2}\left(x,t\right)}\right)\label{eq:Ge}
\end{equation}

and 
\begin{equation}
G_{h}(x-x',y,z,t)=\frac{1}{\sqrt{\left(2\pi\right)^{3}}\sigma_{h}^{3}\left(x,t\right)}\exp\left(-\frac{\left(x-x'\right)^{2}+y^{2}+z^{2}}{2\sigma_{h}^{2}\left(x,t\right)}\right)\label{eq:Gh}
\end{equation}

If $\sigma_{e}\rightarrow0$ and $\sigma_{h}\rightarrow0$ for $t\rightarrow0$
then the functions $G_{e}$ and $G_{h}$ can be regarded as representations
of the $\delta$ function, thus: 
\begin{equation}
G_{e}(x-x',y,z,t=0)=G_{h}(x-x',y,z,t=0)=\delta(x-x')\delta(y)\delta(z)\, dx\label{eq:Gfrom0}
\end{equation}
 Now, the goal is to describe the time evolution of the functions
$B_{e}$, $B_{h}$ and $G_{e}$, $G_{e}$ (for determination of $G_{e}$,
$G_{e}$ it is sufficient to derive the time evolution of their variances
$\sigma_{e}^{2}$ and $\sigma_{h}^{2}$). In order to do it we define
the one dimensional densities associated with the variable $x$ as:

\begin{equation}
\eta_{e}(x,t)=\intop_{_{-\infty}}^{\:\:\infty}dy\intop_{_{-\infty}}^{\:\:\infty}dz\rho_{e}(\mathbf{r,}t)\label{eq:eta_e}
\end{equation}

\begin{equation}
\eta_{h}(x,t)=\intop_{_{-\infty}}^{\:\:\infty}dy\intop_{_{-\infty}}^{\:\:\infty}dz\rho_{h}(\mathbf{r,}t)\label{eq:eta_h}
\end{equation}
 and we divide the thickness of the detector $d_{Si}$ into $N$ intervals
$\Delta x=\frac{d_{Si}}{N}$. Let us assume that in the interval $\Delta x_{i}$
($i=1,N$) the associated values of $\sigma_{ei}$ and $\sigma_{hi}$
do not change substantially within the radius of a few sigma around
$\Delta x_{i}$, and that a linear approximation can be used for the
functions $B_{e}(x',t)$ and $B_{h}(x',t)$ within $\Delta x_{i}$:

\begin{equation}
B_{e}(x',t)=p_{ei}(t)x'+q_{ei}(t)\label{eq:Be_exp}
\end{equation}

\begin{equation}
B_{h}(x',t)=p_{hi}(t)x'+q_{hi}(t)\label{eq:Bh_exp}
\end{equation}
 With the above assumption for $x$ within an interval $\Delta x_{i}$
one can approximate the densities:

\begin{equation}
\eta_{e}(x,t)\backsimeq\intop_{_{-\infty}}^{\:\:\infty}dx'\intop_{_{-\infty}}^{\:\:\infty}dy\intop_{_{-\infty}}^{\:\:\infty}dz\left(p_{ei}(t)x'+q_{ei}(t)\right)G_{e}(x-x',y,z,t)=p_{ei}(t)x+q_{ei}(t)\label{eta_e_bis}
\end{equation}
 
\begin{equation}
\eta_{h}(x,t)\backsimeq\intop_{_{-\infty}}^{\:\:\infty}dx'\intop_{_{-\infty}}^{\:\:\infty}dy\intop_{_{-\infty}}^{\:\:\infty}dz\left(p_{ei}(t)x'+q_{ei}(t)\right)G_{h}(x-x',y,z,t)=p_{hi}(t)x+q_{hi}(t)\label{eta_h_bis}
\end{equation}
 which means that, in practice, one can use the same coefficients
for linear expansion of both, the densities $\eta_{e},$ $\eta_{h}$
and of the functions $B_{e}$, $B_{h}$.

We introduce also the $x_{e0}(t)$ and $x_{eN}(t)$ coordinates, which
denote the beginning and end of the $B_{e}(x,t)$ distributions for
electrons. Similar coordinates $x_{h0}(t)$ and $x_{hN}(t)$ are introduced
for holes (see right-upper panel on Fig. 2).

\subsection{Electric field determination}

In order to determine the drift velocity associated with the centers
of Gaussians $G_{e}$, $G_{h}$ we have to calculate the respective
effective electric field acting on the carriers described by above
distributions. Such a field is determined by a static voltage applied
to the detector electrodes and by the Coulomb interaction between
the Gaussian charge clouds. The detector static field at position
$x$, considered from the rear side of the detector (order $n-p$
from the point of view of the particle entering the detector, see
e.g. \cite{Parlog_10}), is given as:

\begin{equation}
E_{stat}(x)=\frac{2V_{d}x}{d_{Si}^{2}}+\frac{V-V_{d}}{d_{Si}}\label{eq:Ex-stat}
\end{equation}
 where the bias voltage $V$ is assumed to be higher than the depletion
voltage $V_{d}$ which for the bulk concentration of donors $N_{D}$
and permittivity $\varepsilon=\varepsilon_{r}\varepsilon_{0}$ reads
as: 
\begin{equation}
V_{d}=\frac{eN_{D}d^{2}}{2\varepsilon_{r}\varepsilon_{0}}\label{eq:Vdep}
\end{equation}

In order to find the modification of the electric field caused by
the generated plasma, let us consider two Gaussians describing the
distribution of the charges $Z_{1}$, $Z_{2}$, centered at a relative
distance $r_{12}=$$\left|\mathbf{r}_{1}-\mathbf{r}_{2}\right|$ and
characterized by variances $\sigma_{1}^{2}$, $\sigma_{2}^{2}$ respectively.
The mutual interaction potential of the clouds can then be expressed
in the form which can be often found in quantum molecular dynamics
calculations (see e.g. \cite{Papa_01}) 
\[
v(r_{01},\, r_{02},\,\sigma_{1},\,\sigma_{2})=\frac{e^{2}Z_{1}Z_{2}}{\left(2\pi\sigma_{1}\sigma_{2}\right)^{3}}\iint\frac{\exp\left(\frac{-\left(\mathbf{r}_{1}-\mathbf{r}_{01}\right)^{2}}{2\sigma_{1}^{2}}\right)\exp\left(\frac{-\left(\mathbf{r}_{2}-\mathbf{r}_{02}\right)^{2}}{2\sigma_{2}^{2}}\right)}{\left|\mathbf{r}_{1}-\mathbf{r}_{2}\right|}d\mathbf{^{3}r}_{1}d\mathbf{^{3}r}_{2}=
\]
 
\begin{equation}
=e^{2}Z_{1}Z_{2}\frac{\mathrm{erf}\left(\frac{r_{12}}{\sqrt{2}\sigma}\right)}{r_{12}}\label{eq:vc}
\end{equation}
 where $\sigma=\sqrt{\sigma_{1}^{2}+\sigma_{2}^{2}}$.

Let us now assume that the intervals $\Delta x_{i}$ are small enough
to enable the linear approximation for the charge densities $\rho_{e},$
$\rho_{h}$ and the functions $B_{e}$ ,$B_{h}$ with the use of the
coefficients $p$ and $q$. For simplicity, we introduce variables
$p$ , $q$ in lieu of the $p_{ei}(t)$, $p_{hi}(t)$ and $q_{ei}(t)$,
$q{}_{hi}(t)$. If we denote the endpoints of the $\Delta x_{i}$
interval by $c$ and $d$ then for a Gaussian centered at a point
$a$ and representing the charge $Z_{a}$, its interaction with the
charge located in the interval $\Delta x_{i}=(c,\, d)$ given by (\ref{eta_e_bis})
and (\ref{eta_h_bis}), can be formulated as:

\begin{equation}
V_{C}\left(a,c,d,p,q,\mbox{\ensuremath{\sigma_{a}},\ensuremath{\sigma_{i}}}\right)=\frac{e^{2}}{\varepsilon}Z_{a}{\displaystyle \int_{c}^{d}}dx\,\left(px+q\right)\frac{\mathrm{erf}\left(\frac{\left|x-a\right|}{\sqrt{2}\sigma_{s}}\right)}{\left|x-a\right|}\label{eq:V}
\end{equation}
 where the $\sigma_{a}$ and \foreignlanguage{american}{$\sigma_{i}$}
above denote the standard deviations of the Gaussian describing the
charge $Z_{a}$ and of the Gaussian from the interval $\Delta x_{i}$,
respectively, and $\sigma_{s}=\sqrt{\sigma_{a}^{2}+\sigma_{i}^{2}}$.

The above form allows us to describe the respective effective electric
field $E_{x}$ acting on the Gaussian located at a point $a$ as 
\begin{equation}
E_{x}\left(a,c,d,p,q,\mbox{\ensuremath{\sigma_{a}},\ensuremath{\sigma_{i}}}\right)=-\frac{1}{Z_{a}}\frac{\partial V_{C}}{\partial a}=-\frac{e^{2}}{\varepsilon}{\displaystyle \int_{c}^{d}}dx\,\left(px+q\right)\frac{\partial}{\partial a}\frac{\mathrm{erf}\left(\frac{\left|x-a\right|}{\sqrt{2}\sigma_{s}}\right)}{\left|x-a\right|}\label{eq:Ex-int}
\end{equation}
 Since $\frac{\partial}{\partial a}$ = $-\frac{\partial}{\partial x}$,
the above formula can be expressed as:

\[
E_{x}\left(a,c,d,p,q,\mbox{\ensuremath{\sigma_{a}},\ensuremath{\sigma_{i}}}\right)=\frac{e^{2}}{\varepsilon}{\displaystyle \int_{c}^{d}}dx\,\left(px+q\right)\frac{\partial}{\partial x}\frac{\mathrm{erf}\left(\frac{\left|x-a\right|}{\sqrt{2}\sigma_{s}}\right)}{\left|x-a\right|}=
\]

\begin{equation}
=\frac{e^{2}}{\varepsilon}q\left(\frac{\mathrm{erf}\left(\frac{\left|d-a\right|}{\sqrt{2}\sigma_{s}}\right)}{\left|d-a\right|}-\frac{\mathrm{erf}\left(\frac{\left|c-a\right|}{\sqrt{2}\sigma_{s}}\right)}{\left|c-a\right|}\right)+\frac{e^{2}}{\varepsilon}{\displaystyle p\int_{c}^{d}}dx\, x\frac{\partial}{\partial x}\frac{\mathrm{erf}\left(\frac{\left|x-a\right|}{\sqrt{2}\sigma_{s}}\right)}{\left|x-a\right|}\label{Ex_int_1}
\end{equation}

After integrating by parts one obtains:

\[
E_{x}=\frac{e^{2}}{\varepsilon}q\left(\frac{\mathrm{erf}\left(\frac{\left|d-a\right|}{\sqrt{2}\sigma_{s}}\right)}{\left|d-a\right|}-\frac{\mathrm{erf}\left(\frac{\left|c-a\right|}{\sqrt{2}\sigma_{s}}\right)}{\left|c-a\right|}\right)+\frac{e^{2}}{\varepsilon}p\left(d\frac{\mathrm{erf}\left(\frac{\left|d-a\right|}{\sqrt{2}\sigma_{s}}\right)}{\left|d-a\right|}-c\frac{\mathrm{erf}\left(\frac{\left|c-a\right|}{\sqrt{2}\sigma_{s}}\right)}{\left|c-a\right|}\right)+
\]
 
\begin{equation}
{\displaystyle -\frac{e^{2}}{\varepsilon}p\int_{c}^{d}}dx\,\frac{\mathrm{erf}\left(\frac{\left|x-a\right|}{\sqrt{2}\sigma_{s}}\right)}{\left|x-a\right|}\label{eq:Ex_int_2}
\end{equation}
 The last integral can be easily evaluated by expanding the error
function.

The total effective electric field modified due to the presence of
plasma can be obtained by summing up contributions of charges located
in all intervals $(c_{i},d_{i})$ and of their mirror charges induced
in the detector electrodes (see Fig. 1).

In order to determine the time evolution of the charge distribution
one has to know, in addition, the generalized force associated with
the $\sigma_{a}$ variable. Similarly as for the effective electric
field, the net value of the force acting in the $\sigma_{a}$ direction
can be obtained by summing over all ingredients associated with the
charge distribution. This leads to the following expression for the
interaction with the charge located in the $\Delta x_{i}$ interval:
\[
F_{\sigma_{a},a,cd}=-\frac{\partial V_{C}}{\partial\sigma_{a}}=-\frac{e^{2}}{\varepsilon}Z_{a}{\displaystyle \int_{c}^{d}}dx\,\left(px+q\right)\frac{\partial}{\partial\sigma_{a}}\frac{\mathrm{erf}\left(\frac{\left|x-a\right|}{\sqrt{2}\sigma_{s}}\right)}{\left|x-a\right|}=
\]
 
\[
=-Z_{a}\frac{e^{2}p\sigma_{a}}{\sqrt{\frac{\pi}{2}}\varepsilon\sigma_{s}}\left(\exp\left(-\frac{\left(a-d\right)^{2}}{2\sigma_{s}^{2}}\right)-\exp\left(-\frac{\left(a-c\right)^{2}}{2\sigma_{s}^{2}}\right)\right)+
\]
 
\begin{equation}
+Z_{a}\frac{e^{2}\sigma_{a}(ap+q)}{\varepsilon\sigma_{s}^{2}}\left(\mathrm{erf}\left(\frac{\left(a-d\right)}{\sqrt{2}\sigma_{s}}\right)-\mathrm{erf}\left(\frac{\left(a-c\right)}{\sqrt{2}\sigma_{s}}\right)\right)\label{eq:F_sig}
\end{equation}

Now we are in position to calculate the time evolution of the charge
generated in the detector. This process is determined by the drift
and by the diffusion of the interacting clouds of electrons and holes.
We will consider the evolution of the centroids and of the variances
of Gaussians representing a fraction of the charge distribution located
in the middle of the intervals $\Delta x_{i}$ and at the start- and
end-points of the distributions of electrons and holes (points $x_{e0}(t)$,
$x_{eN}(t)$ and $x_{h0}(t)$, $x_{hN}(t)$) .

\subsection{Evolution of the function \emph{B}}

In the present subsection we will describe the numerical method used
to determine the time evolution of the ionization clouds. In the following
we assume that the evolution of the functions $B_{e}$, $B_{h}$ is
determined by the time evolutions of the coefficients $p_{ei}$, $q_{ei}$
and $p_{hi}$, $q_{hi}$. In order to find how the expansion coefficients
$p_{ei}$, $q_{ei}$ and $p_{hi}$ , $q_{hi}$ propagate in time,
we have to investigate the time evolution of the functions $\eta_{e}(x,t)$
and $\eta_{h}(x,t)$.

Below we consider the formulas for electrons only, keeping in mind
that the formulas for holes are analogical.

As we will show later

\begin{equation}
\frac{\partial}{\partial t}\eta_{e}(x,t)=-\frac{\partial}{\partial x}\left(\eta_{e}(x,t)v_{xe}(x,t)\right)\label{eq:dif_eta}
\end{equation}
 thus, the differential $d\eta_{e}(x,t)$ can be written as

\begin{equation}
d\eta_{e}(x,t)=-\left(v_{xei}(x,t)\frac{\partial}{\partial x}\eta_{e}(x,t)+\eta_{e}(x,t)\frac{\partial}{\partial x}v_{xe}(x,t)\right)dt\label{eq:inc_eta}
\end{equation}

If the $x_{i}$ denotes the center of the interval $\Delta x_{i}$
and $q_{ei}$ and $p_{ei}$ are the coefficients of linear expansion

\begin{equation}
\eta_{e}(x,t)=p_{ei}(t)(x-x_{i})+q_{ei}\label{eq:exp_eta}
\end{equation}
 and, if one denotes the average velocity and the average linear density
associated with the interval $\Delta x_{i}$ by $v_{xei}(t)$ and
$\eta_{xei}(t)$, respectively, then:

\begin{equation}
d\eta_{ei}(t)=-\left(v_{xei}(t)p_{ei}(t)+\eta_{ei}(t)\varphi_{xei}(t)\right)dt\label{eq:increment_eta}
\end{equation}
 Here, $\varphi_{xei}(t)$ is the differential coefficient of $v_{xei}(x,t)$
at the point $x_{i}$. As one can see, in order to calculate the above
increment, we have to trace the time dependent tables of $\eta_{ei}(t)$,
\foreignlanguage{american}{$v_{xei}(t)$}. The tables of $p_{ei}(t)$,
$q_{ei}(t)$ and $\varphi_{xei}(t)$ are obtained by fitting the smooth
curves to the distributions $\eta_{ei}(t)$, \foreignlanguage{american}{$v_{xei}(t)$}
in every time step. For $t=0$ the $\eta_{ei}(t=0)$ is given by the
Bragg curve. In order to make use of the formula (\ref{eq:increment_eta})
we need to construct the respective tables for velocities. Knowing
the effective electric field for electrons in the $x$ direction,
$E_{xei}$, one can assume that the respective average velocity of
the center of Gaussian located at a point $x_{i}$ is proportional
to the strength of the field:

\begin{equation}
v_{xei}=\mu_{xe}E_{xei}\label{eq:velo_x}
\end{equation}

where $\mu_{xe}$ and $\mu_{xh}$ are the electron and hole mobilities,
respectively.

Knowing the drift velocity, one can calculate the evolution of the
charge deposited in every interval $\Delta x_{i}$ including the edge
intervals with variable ends $x_{e0}(t)$, $x_{eN}(t)$ and $x_{h0}(t)$,
$x_{hN}(t)$.

\subsection{Charge propagation in the perpendicular direction }

The diffusion and transport processes in the electric field influence
also the widths of the charge distributions located in every $\Delta x_{i}$
interval. Extending the above reasoning we can assume that the velocity,
$v_{\sigma ei}$, describing the rate of the standard deviation expansion
in perpendicular direction has three components: 
\begin{equation}
v_{\sigma ei}=v_{\sigma ei}^{E}+v_{\sigma ei}^{D}+v_{\sigma ei}^{T}\label{eq:velo_sig}
\end{equation}

The first term, $v_{\sigma ei}^{E}$, results from the field described
by (\ref{eq:F_sig}). In analogy to the charge drift in $x$ direction,
one can assume that this component is proportional to the field acting
on the charge $Z_{a}$ associated with the Gaussian with a standard
deviation $\sigma_{a}$ 
\begin{equation}
E_{\sigma ei}=\frac{F_{\sigma_{a}}}{Z_{a}}\label{eq:E_sig}
\end{equation}
 where $F_{\sigma_{a}}$ is the net force given by interaction (\ref{eq:F_sig}).
Thus, the respective velocity can be expressed as

\begin{equation}
v_{\sigma ei}^{E}=\mu_{\sigma e}E_{\sigma ei}\label{eq:v_E_sig}
\end{equation}
 where, the $\mu_{\sigma e}$ parameter is the only free parameter
of the model. It seems, however, that it can be determined theoretically
in the future. A similar parameter for description of the hole propagation
can be calculated assuming the following proportion:

\begin{equation}
\frac{\mu_{\sigma e}}{\mu_{\sigma h}}=\frac{\mu_{xe}}{\mu_{xh}}\label{eq:mi_mi}
\end{equation}
 The velocity $v_{\sigma ei}^{D}$ follows from the solution of the
second Fick's law for diffusions of Gaussian density distributions:
\begin{equation}
v_{\sigma ei}^{D}=\frac{\partial\sigma_{ei}}{\partial t}\mid_{v_{\sigma ei}^{E}=0,v_{\sigma ei}^{T}=0}=\frac{D_{e}}{\sigma_{ei}}\label{eq:v_D_sig}
\end{equation}
 where $D_{e}$ is the diffusion coefficient for electrons.

Another process which affects, on average, the widths of the respective
Gaussian distributions used to describe the charge located in the
interval $\Delta x_{i}$, is related to the transport of the carriers.
In order to describe this process we consider an increase of the variance,
$\sigma^{2}$, of the Gaussian distribution of the charge of $Z$
particles contained in an interval $\Delta x$, with linear density
$\eta(x)=\frac{Z}{\Delta x}$. Let the $\sum{\displaystyle r_{i}^{2}}$
denote the sum of squares of deviations of particle positions from
the average. In this consideration we neglect the influence of the
diffusion and of mutual interactions of clouds on the propagation
of the variance.

The change of $\sigma^{2}\equiv\frac{\sum{\displaystyle r_{i}^{2}}}{Z}$
, resulting from the flow of particles into and out of the cell $\Delta x$
(with the net value of $dZ$) is, in general, equal to 
\begin{equation}
d\sigma^{2}=\frac{d\left(\sum{\displaystyle r_{i}^{2}}\right)}{Z}-\frac{\sigma^{2}(x)dZ}{Z}\label{eq:dsigma_2}
\end{equation}
 As one can see, in order to find the increment $d\sigma^{2}$ we
have to find the increments $dZ$ and $d\left(\sum{\displaystyle r_{i}^{2}}\right)$.

Let us begin with the description of $d\left(\sum{\displaystyle r_{i}^{2}}\right)$.
If the accretion of particles in an interval $\Delta x$ across the
points $x_{1}=x-\Delta x/2$ and $x_{2}=x+\Delta x/2$ is denoted
by $dZ_{1}$and $dZ_{2}$, respectively, and the variances at these
points are denoted by $\sigma_{1}^{2}=\sigma^{2}(x-\Delta x/2$) and
$\sigma_{2}^{2}=\sigma^{2}(x+\Delta x/2$), respectively, then the
increment of the sum $\sum{\displaystyle r_{i}^{2}}$ can be determined
as:

\begin{equation}
d\left(\sum{\displaystyle r_{i}^{2}}\right)=\sigma_{1}^{2}dZ_{1}-\sigma_{2}^{2}dZ_{1}\label{eq:dsr}
\end{equation}
 Denoting the velocities of particles at points $x_{1}$ and $x_{2}$
by $v(x-\Delta x/2)$ and $v(x+\Delta x/2)$, respectively, the accretions
~$dZ_{1}$ and $dZ_{2}$ can be determined as 
\begin{equation}
dZ_{1}=\eta(x-\Delta x/2)\, v(x-\Delta x/2)dt\label{eq:dz1_1}
\end{equation}
 
\begin{equation}
dZ_{2}=\eta(x+\Delta x/2)\, v(x+\Delta x/2)dt\label{eq:dz2_1}
\end{equation}

Now one can calculate the increment $d\left(\sum{\displaystyle r_{i}^{2}}\right)$
as

\[
d\left(\sum{\displaystyle r_{i}^{2}}\right)=-dt\cdot\Delta x\cdot
\]

\begin{equation}
\cdot\left[\frac{\sigma^{2}(x+\Delta x/2)\eta(x+\Delta x/2)v(x+\Delta x/2)-\sigma^{2}(x-\Delta x/2)\eta(x-\Delta x/2)v(x-\Delta x/2)}{\Delta x}\right]\label{eq:dsr_2}
\end{equation}
 The expression in square brackets tends to the partial derivate $\frac{\partial(\sigma^{2}(x)\eta(x)v(x))}{\partial x}$
for $\Delta x\rightarrow0$.

Similarly, the increment $dZ=dZ_{1}-dZ_{2}$ can be written as

\begin{equation}
dZ=-\Delta x\left[\frac{\eta(x+\Delta x/2)v(x+\Delta x/2)-\eta(x-\Delta x/2)v(x-\Delta x/2)}{\Delta x}\right]dt\label{eq:dZ}
\end{equation}
 and again, in the limit of $\Delta x\rightarrow0$ the expression
in square brackets approaches to $\frac{\partial(\eta(x)v(x))}{\partial x}$,
which has already been used in (\ref{eq:dif_eta}).

Taking the above into account, setting $Z=\eta(x)\Delta x$ and taking
the $\Delta x\rightarrow0$ limit, eq. (\ref{eq:dsigma_2}) can be
transformed into:

\begin{equation}
d\sigma^{2}(x)=\left[\frac{-\partial\left(\sigma^{2}(x)\eta(x)v(x)\right)}{\partial x}+\frac{\sigma^{2}(x)\partial(\eta(x)v(x))}{\partial x}\right]\frac{dt}{\eta(x)}=\frac{-v(x)\partial\left(\sigma^{2}(x)\right)}{\partial x}dt\label{eq:inc_s2}
\end{equation}
 what gives

\begin{equation}
\frac{\partial\sigma^{2}}{\partial t}=-v\frac{\partial\sigma^{2}}{\partial x}\;\Rightarrow\;\frac{\partial\sigma}{\partial t}=-v\frac{\partial\sigma}{\partial x}\label{eq:dif_s2}
\end{equation}
 Finally for $v_{\sigma ei}^{E}=0$ and $v_{\sigma ei}^{D}=0$ one
can write 
\begin{equation}
v_{\sigma ei}^{T}=\frac{\partial\sigma_{ei}}{\partial t}\mid_{v_{\sigma ei}^{E}=0,v_{\sigma ei}^{D}=0}=-v_{xei}\frac{\partial\sigma_{ei}}{\partial x}\label{eq:v_T_sig}
\end{equation}

In order to use the above formula we trace the changes of the vector
of$\sigma_{ei}$ values as a function of the position index, $i$.
Knowledge of velocities $v_{\sigma ei}^{E}$, $v_{\sigma ei}^{D}$
and $v_{\sigma ei}^{T}$ allows to calculate the propagation of the
width of the distribution of electrons. The formulas for holes are
analogical.

\section{First prediction of the model and comparison with the experimental
data}

For the first comparison of the model prediction with the experimental
data we choose the data for $^{12}C$ ion which have already been
used in \cite{Parlog_10}. This gives also the opportunity to compare
the present model predictions with those obtained in a more phenomenological
approach. For the measurement the neutron transmutation doped (n-TD)
silicon detector \cite{Ammon_92} was used. This n-type bulk and extremely
thin p-type zone has a thickness $d_{Si}=310\;\mu m$. The energy
measurement was performed using the charge output, while the current
pulses were measured using the current output of the same preamplifier,
described in \cite{Harmita_04}. This paper presents also in detail
the experimental setup and conditions used for the$^{12}C$ ions (and
LCP).

Before describing the induced current pulse, we focus first on the
propagation of the electric field and the propagation of the electron
and hole densities in parallel and perpendicular directions. The evolution
of these observables is important for understanding the mechanism
of the plasma delay process. In the following, we consider an $^{12}C$
ion impinging on the n-type rear side of the silicon detector. This,
so called {}``rear-mount'', gives quite different shapes as compared
to the {}``standard mount'', and these pulse shapes are much better
suited for the PSD technique \cite{Harmita_04}.

For actual calculations it is necessary to set some physical coefficients
describing the electric field propagation, as well as coefficients
describing the drift and diffusion process in silicon. In table 1
we collect values of these parameters: 
\[
\]

\begin{tabular}{|c|c|c|c|}
\hline 
feature  & symbol  & value  & remarks\tabularnewline
\hline 
\hline 
energy per e\textendash{}h pair creation  & $w$  & 3.6 $\left[\frac{eV}{pair}\right]$  & material constant\tabularnewline
\hline 
silicon dielectric, permittivity  & $\varepsilon_{r}$  & 11.7  & material constant\tabularnewline
\hline 
electrons mobility  & $\mu_{xe}$  & 135 $\left[\frac{\mu m^{2}}{Vns}\right]$  & material constant\tabularnewline
\hline 
holes mobility  & $\mu_{xh}$  & 47.5 $\left[\frac{\mu m^{2}}{Vns}\right]$  & material constant\tabularnewline
\hline 
electrons variance mobility  & $\mu_{\sigma e}$  & 2 $\left[\frac{\mu m^{2}}{Vns}\right]$  & free parameter\tabularnewline
\hline 
holes variance mobility  & $\mu_{\sigma h}$  & $\mu_{\sigma h}=\mu_{\sigma e}\frac{\mu_{xh}}{\mu_{xe}}$  & model assumption\tabularnewline
\hline 
diffusion coefficient for electrons  & $D_{e}$  & 3.49$\left[\frac{\mu m^{2}}{ns}\right]$  & material constant\tabularnewline
\hline 
diffusion coefficient for holes  & $D_{h}$  & 1.228$\left[\frac{\mu m^{2}}{ns}\right]$  & material constant\tabularnewline
\hline 
\end{tabular}

\[
\]

As already mentioned, the electric field propagation results from
the static detector bias and from the generated charge density propagation.
At the starting point, when the generated electrons and holes are
almost exactly at the same positions, the electric field is still
equal to the external one (\ref{eq:Ex-stat}). This field, for t=0,
is denoted in Fig. 2 by a doted line. After this initial moment the
static field causes the shift of the electron and hole distributions
and therefore in the next moment some of the carriers are moved outside
of the overlap region. Next, the variance of the Gaussian partial
density distribution associated with these carriers begins to grow,
due to the non-compensated electric field in perpendicular direction.
This effect, which is displayed in Fig. 3, causes breaking of the
initial bonds between the electrons and holes. As a result, the considered
electrons and holes start leaking slowly from the overlap region and
begin moving in opposite directions. At the same time, the increasing
shift between electrons and holes leads to significant reduction of
the electric field in the interaction region. This scenario is well
associated with the postulates of the phenomenological model of ref.
\cite{Parlog_10}.

The evolution of the charge density, the effective electric field
and the variances of the Gaussian partial densities are presented
in Figs 2-5 in which the blue lines represent the dependences for
the electrons while the red ones represent the holes. Fig. 2 shows
the evolution of the linear electron and hole densities. One can see
that, during the first 20-30 ns, the electrons and holes remain bound
in the region close to the detector surface. After that time, one
can observe the electric field restitution practically in the whole
detector area. The time behavior of the electric field is presented
in Fig. 3. In order to save the calculation time, the field is calculated
only at points where the density of the particles is not equal to
zero. Fig. 4 presents the evolution of the width of the charge distributions.
We can notice that in the overlap region the effective electric field
is reduced to very low value. As one can see, the evolutions of the
electric field and of the charge densities (in parallel and perpendicular
directions) are strongly correlated.

Up to now, the mutual Coulomb interactions between the charge clouds
have been taken into account. In order to see the importance of these
mutual interactions, they have been neglected in the charge density
evolutions presented in Fig 5. As one can see, in this case, the collection
time becomes about tree times shorter, due to the lack of binding
between the electrons and holes. Knowledge of the charge propagation,
by using of the Ramo-Shockley theorem \cite{Shockley_38,Ramo_39,Hunsuk_91},
allowed as to determine the current pulse time dependence.

In order to obtain a rough estimate of the pulse shape, we describe
the partial current associated with the Gaussian cloud in approximate
way. For simplification, using Ramo-Shockley theorem, we replace the
Gaussian charge distribution by a point-like one. Such an approach
neglects effects associated with the charge diffuseness, particularly
for clouds moving closely to the detector electrodes.

The result of such a calculation is presented in fig 6. The total
current pulse is denoted by black solid line while the electron and
hole contributions are represented by the blue and red ones, respectively.

For comparison of the calculated pulse with the experimental one the
primary pulses from Fig. 6. have been corrected (see {[}3{]}) for
the preamplifier's response. The results are shown in Fig. 7. We have
to stress that in the present calculations we did not search for the
best value of the $\mu_{\sigma e}$ parameter. We also did not consider
some quite complicated factors, specified below, which could affect
the obtained results and which will be a subject of the forthcoming
paper:

i) precision of the Energy-Range tables (average accuracy of about
10\%, see \cite{Zigler_85}),

ii) diffuseness of the Gaussian clouds and its presence in the application
of the Ramo-Shockley theorem,

iii) dead layers of the detector and their effect on the measured
energy (as one can see on Fig. 8, for 80 MeV 12C the collection time
is very sensitive on the ion energy),

iv) accuracy of the active detector thickness and of the electric
field determination,

v) accuracy of the preamplifier response description.

In order to demonstrate that the present model is able to describe
correctly the plasma delay effect, in Fig. 8 we present the correlation
between the collection time and the energy loss for $^{10}B$ , $^{12}C$
, and $^{14}N$ ions. We use the collection time rather than the experimentally
preferred rise time, noting that these two observables are strongly
correlated. Fig. 8 shows that the model can reproduce the experimental
trends, especially the characteristic {}``back-bending'' of identification
curves at low energies.

\section{Conclusions}

We have proposed a description of the evolution of charge density
and of the effective electric field by taking into account the mutual
Coulomb interactions between the charge carrier distributions. According
to the present approach the plasma delay effect is associated with
the propagation of the carriers in both directions, perpendicular
and parallel to the primary ionization path.

The duration of the obtained pulse for the 80 MeV $^{12}C$ ion, corresponds
quite well to the one obtained in the physical measurement. Also the
shape of the Energy-Collection Time correlation and its element dependence
are quite well reproduced by the model. Nevertheless, the model still
needs to be confronted with a broader collection of the experimental
data obtained for detectors of various thicknesses and biased by various
voltages.

Once tested on a broader collection of the experimental pulse shapes,
the model will enable the theoretical search for the best identification
method based on the pulse shape analysis. It will also enable the
study of the dependence of the identification quality on the detector
thickness and maybe on some special construction of the detector with
non-linear electric field (obtained by the heterogeneity of doping)
for regions with poor resolution (small ion energy, see Fig. 8). The
presented approach should be also suitable for testing the temperature
dependence (via the respective dependence of the diffusion coefficients).

In order to draw some more quantitative conclusions from the comparison
of the model with the experimental data one has to estimate the uncertainties
related to various possible ingredients, mentioned in the previous
paragraph. Also we have to make an attempt to constrain the single
free parameter $\mu_{\sigma e}$ from the respective classical consideration.

The actual model calculation is quite time consuming. For standard
processor the calculation of one pulse associated with the 80 MeV
$^{12}C$ ion, takes about 5 hours of CPU, thus some code optimization
is still needed.

Acknowledgment:

The author is indebted to J. \L{}ukasik for careful reading of the
manuscript and helpful discussions. Special thanks for H. Hamrita,
for his calculations of the preamplifier response.

Work supported by Polish Ministry of Science and Higher Education
under grant No. DPN/N108/GSI/2009

\section*{Figures}

Figure 1. Allowing the charge induced by plasma in the detector electrodes.
In the present approximation we consider only the nearest mirror reflections.
So the influence of the charge induced in an electrode positioned
at $0$ on the detector electric field, acts as a part of the Gaussian
localized in position $-x$ with the reverse charge and with the same
variance. Similarly, the charge induced in the second electrode is
represented by a respective Gaussian localized in a symmetric point
at $2d_{Si}-x$ .

\includegraphics[scale=0.5]{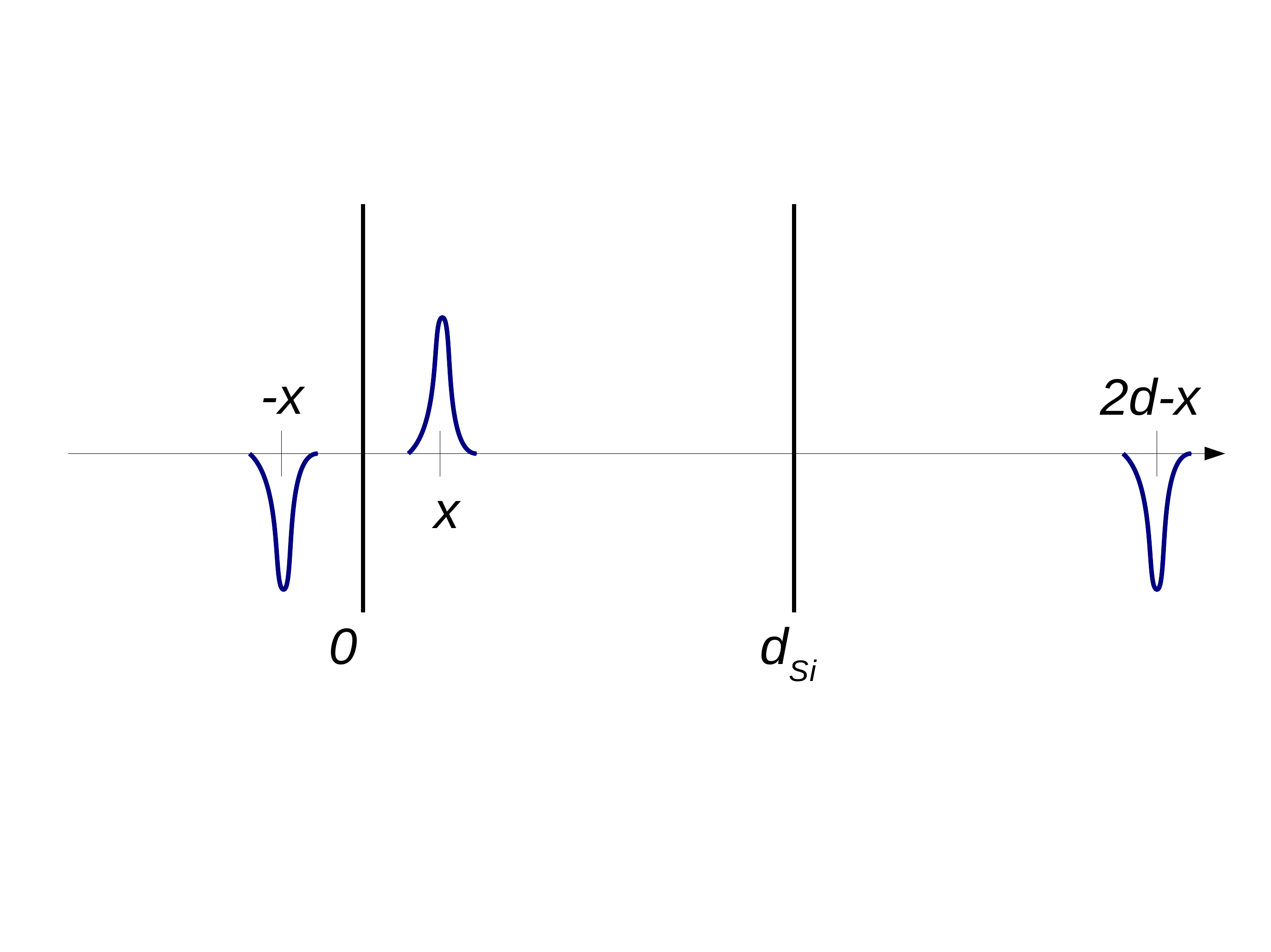}

Figure 2. Propagation of the linear density of the electrons $\eta_{e}$
(blue lines) and the holes $\eta_{h}$ (read lines) due to the ionization
induced by 80 MeV $^{12}C$ ion entering the Si detector from the
rear side.

\includegraphics[scale=0.7]{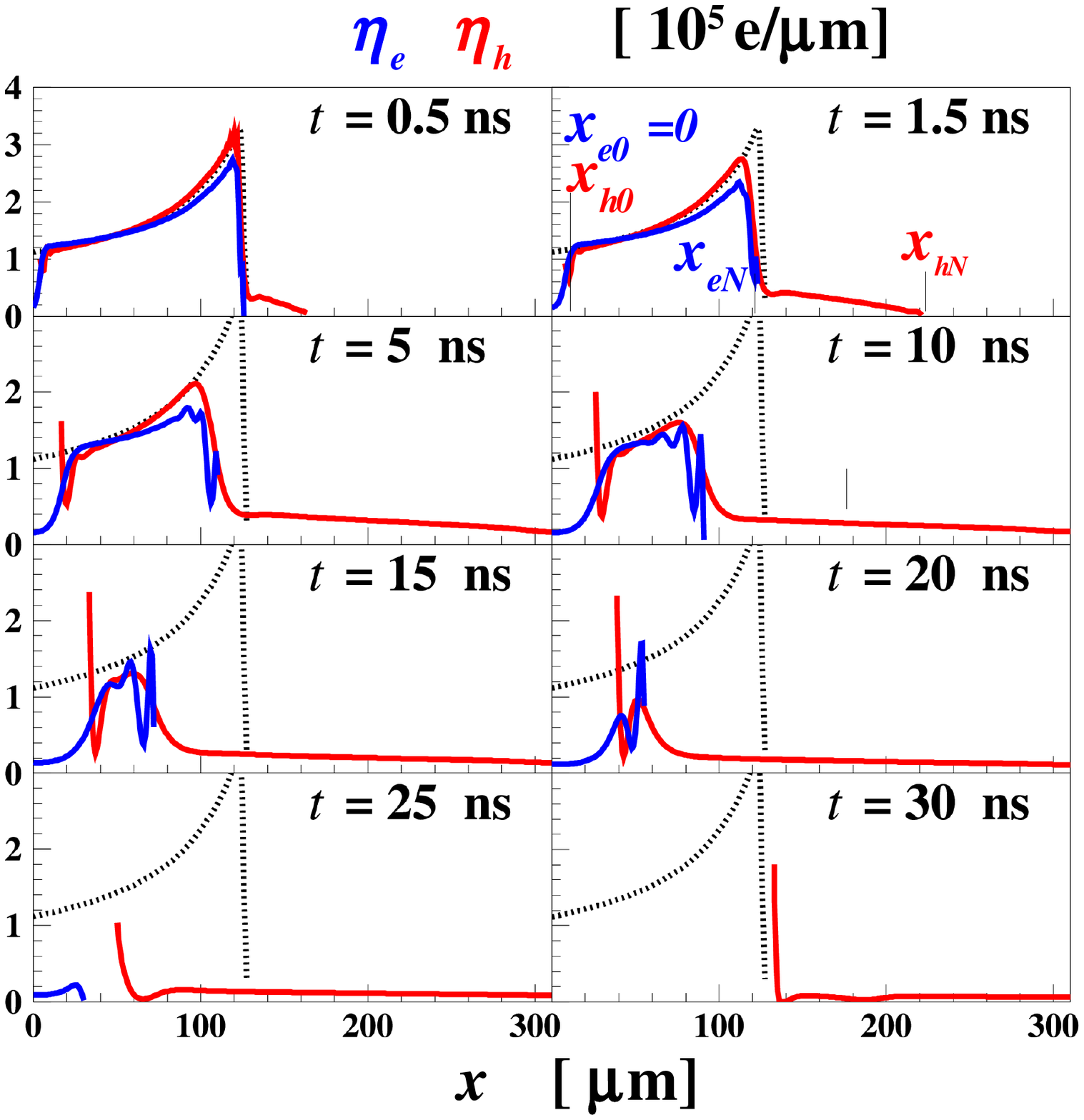}

Figure 3. The effective electric field strength inside the silicon
detector at different moments in time, due to the ionization induced
by 80 MeV $^{12}C$ ion penetrating the detector from the rear side.
The doted line gives the undisturbed electric field for $t=0$. The
field is calculated only at points with coordinate $x$ where the
the density of electrons (blue line) and holes (red line) is not equal
to zero.

\includegraphics[scale=0.7]{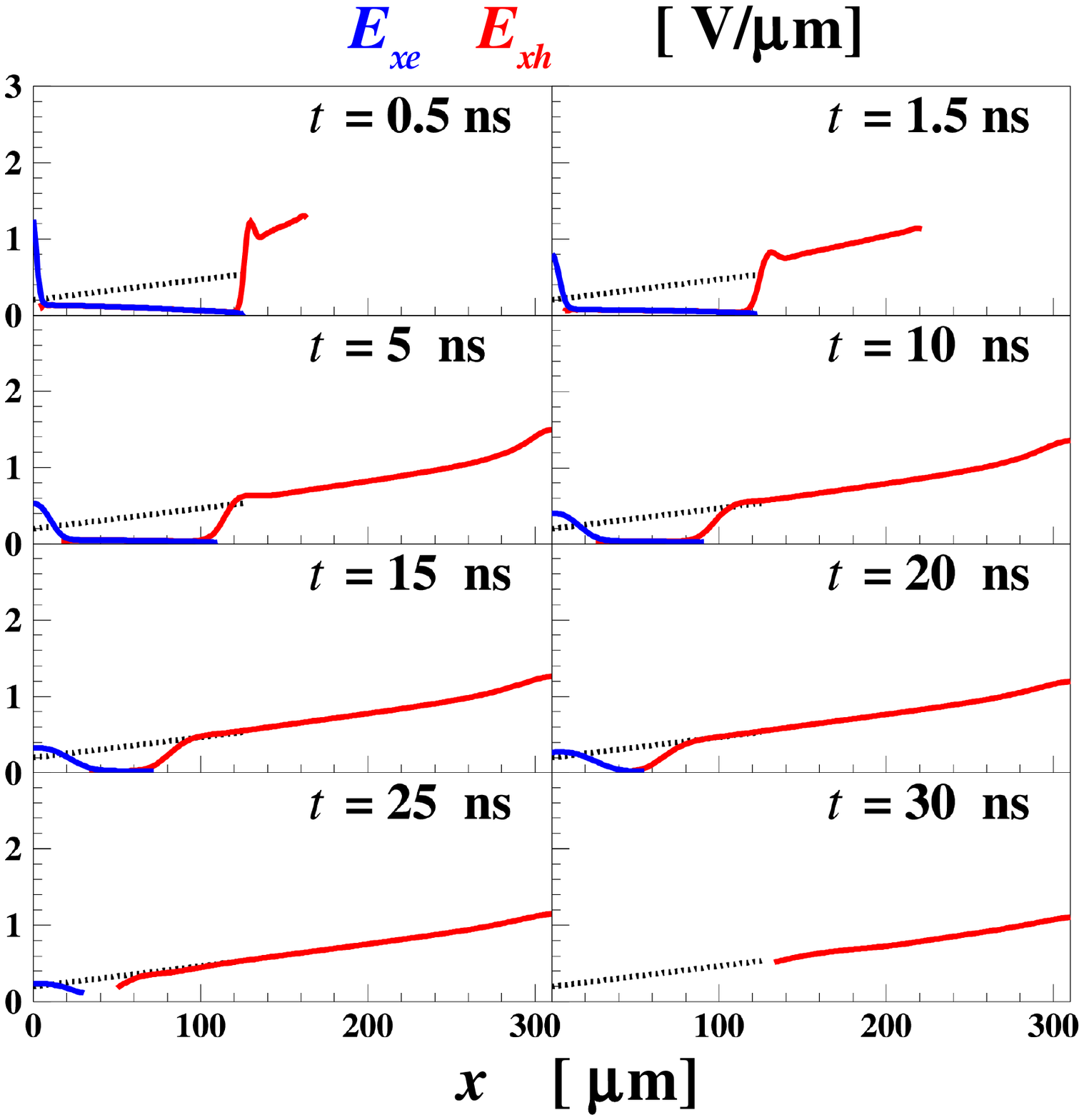}

Figure 4. Time evolution of the width of the charge distribution which
determines the charge propagation in the perpendicular direction.
Standard deviations for electrons are represented by the blue line
and variances for the holes by the red one.

\includegraphics[scale=0.7]{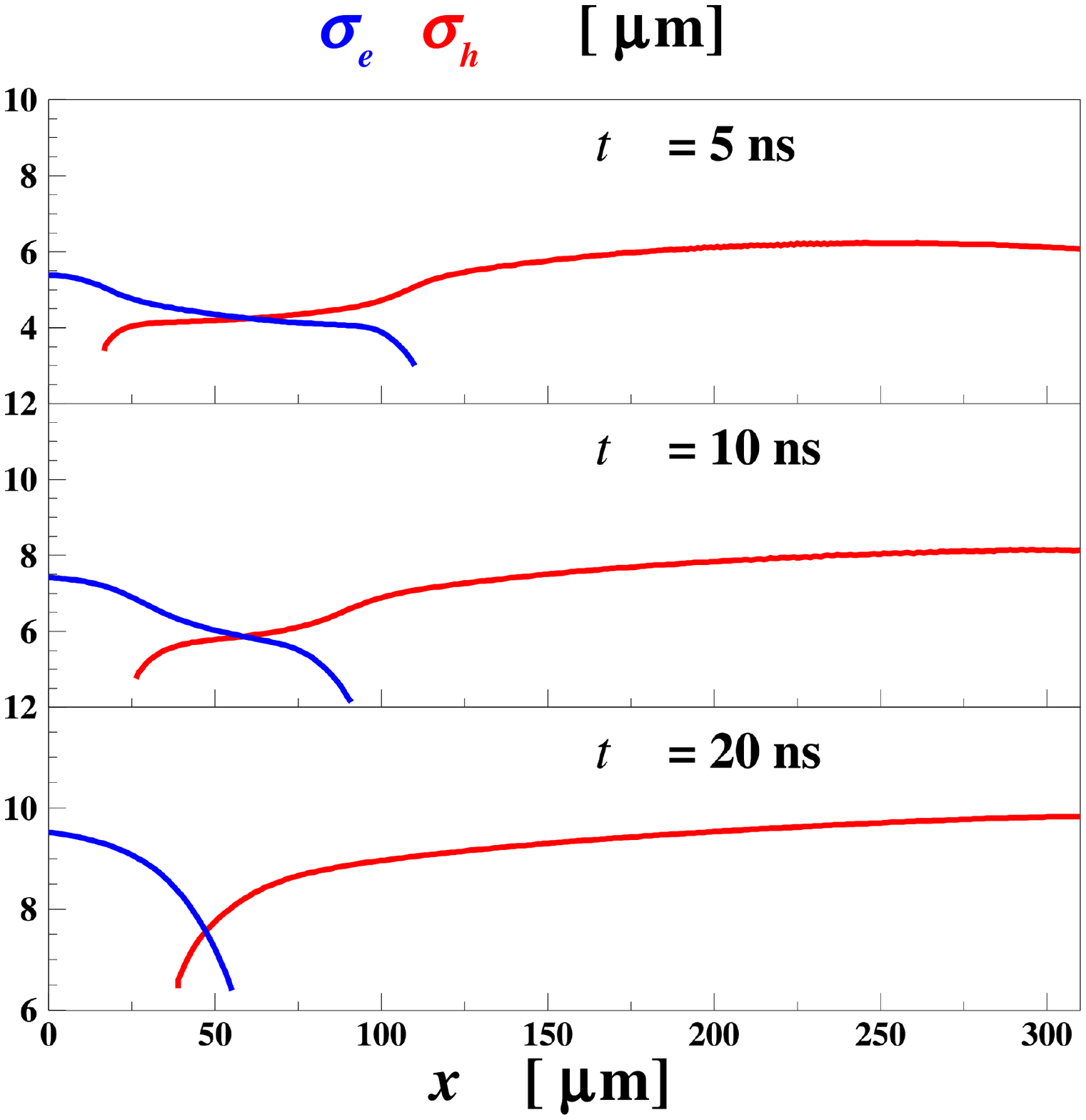}

Figure 5. Same as Fig. 2, but neglecting the mutual Coulomb interactions
between the carrier clouds. Significant difference in the charge collection
time can be observed as compared to the complete case (see Fig. 2).

\includegraphics[scale=0.7]{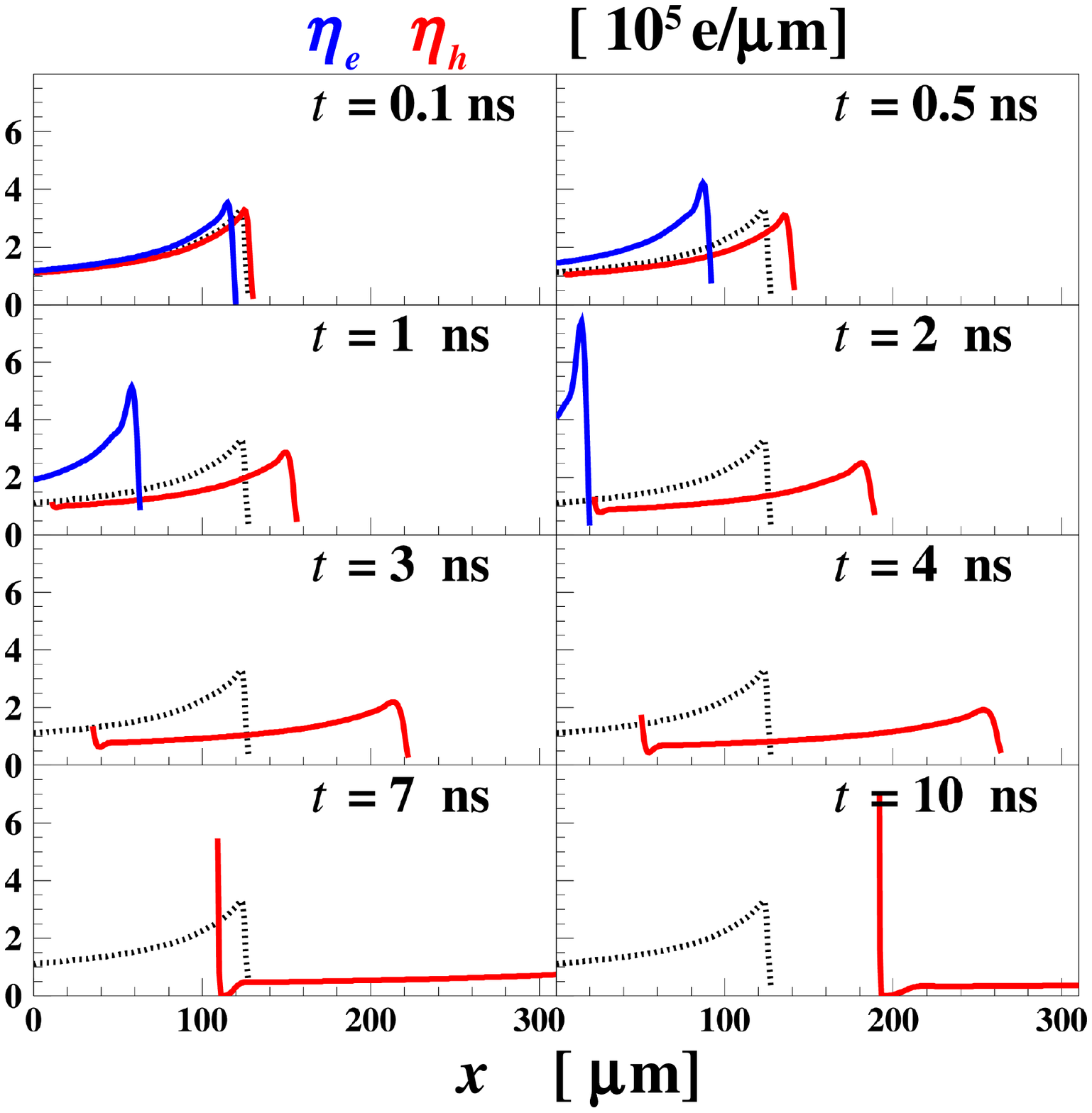}

Figure 6. Model prediction of the current signal induced by an 80
MeV $^{12}C$ ion penetrating the silicon detector from the rear side.
The predictions are not corrected for the preamplifier response. The
solid, black line represents the total signal, the blue line presents
the electron contribution while the red one the hole contribution.
Mean experimental current signal is presented by the dashed line.

\includegraphics[scale=0.7]{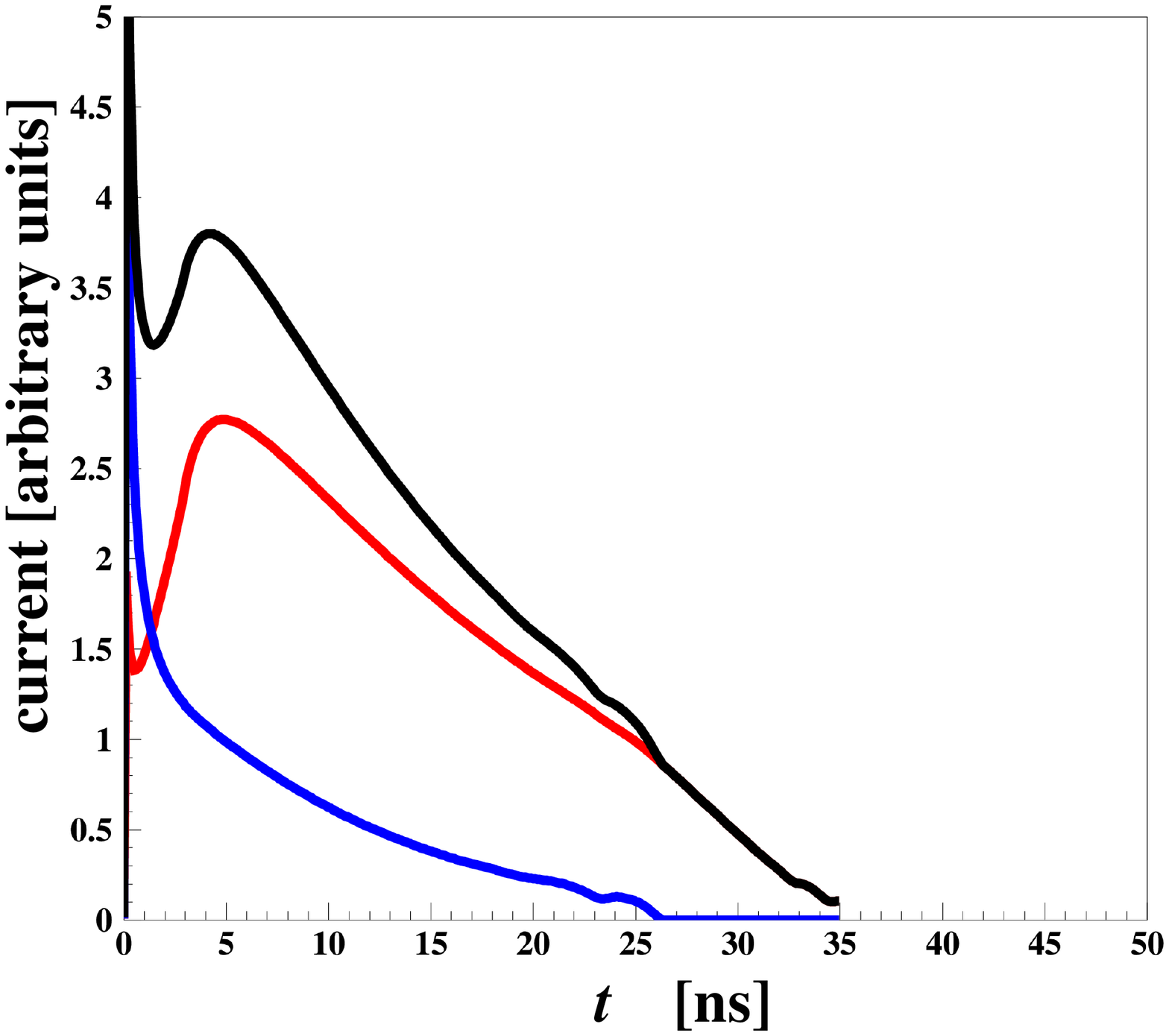}

Figure 7. Same as Fig. 6 but the respective lines have been corrected
for the preamplifier's response.

\includegraphics[scale=0.7]{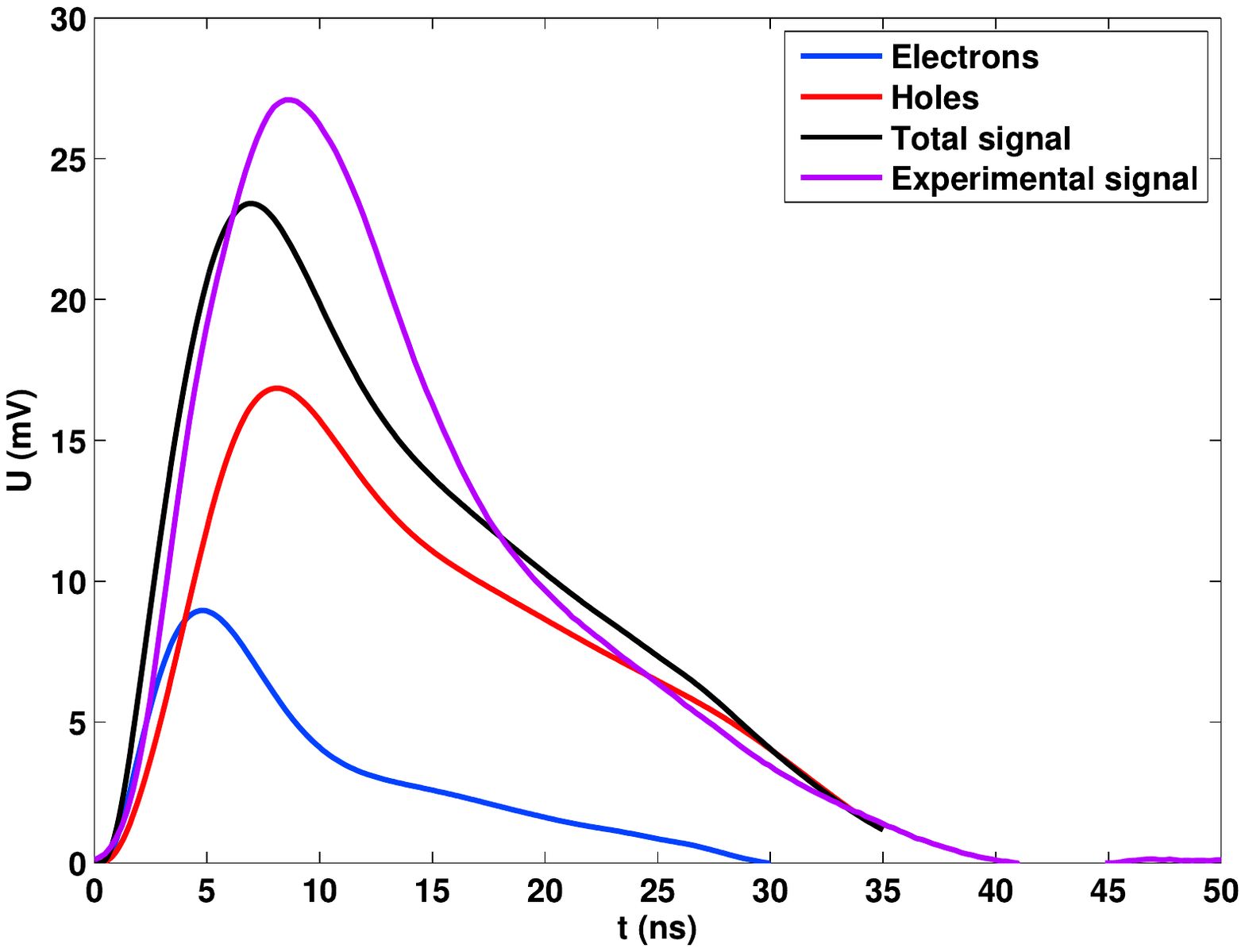}

Figure 8. Model prediction for correlations: Energy vs Collection
Time.

\includegraphics[scale=0.7]{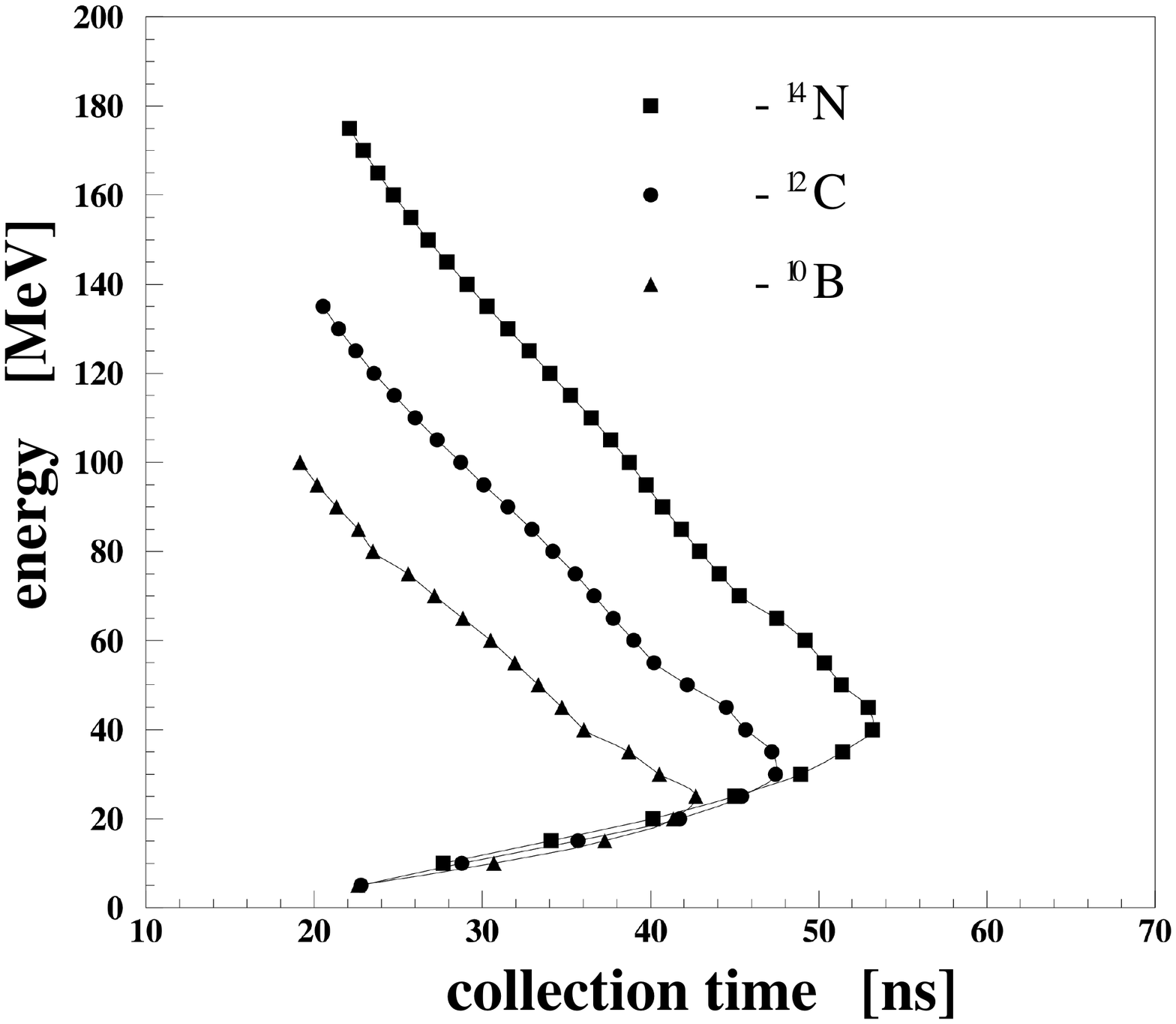}

\begin{thebibliography}{References}
\bibitem{Seibt_73}W. Seibt, et al., Nucl. Instr. and Meth. 113 (1973)
317.

\bibitem{Tove_67}P.A. Tove, W. Seibt, W. Leitz, Nucl. Instr. and
Meth. 51 (1967) 304.

\bibitem{Quaranta_68}A.A. Quaranta, A. Taroni, G. Zanarini, IEEE
Trans. Nucl. Sci. NS-15 (1968) 373.

\bibitem{Neidel_80}H.O. Neidel, H. Henschel, Nucl. Instr. and Meth.
178 (1980) 137.

\bibitem{Bohne_85}W. Bohne, et al., Nucl. Instr. and Meth. A 240
(1985) 145.

\bibitem{England_89}J.B. England, G.M. Field, T.R. Ophel, Nucl. Instr.
and Meth. A 280 (1989) 291.

\bibitem{Harmita_04} H. Hamrita, et al., Nucl. Instr. and Meth. A
531 (2004) 607.

\bibitem{Parlog_10} M. Parlog, et al., Nucl. Instr. and Meth. A 613
(2010) 290.

\bibitem{Zigler_85}J.F. Ziegler, et al., The Stopping and Range of
Ions in Matter (SRIM), Pergamon Press, New York, 1985.

\bibitem{Papa_01} M. Papa et al., Phys. Rev. C 64 024612 2001

\bibitem{Ammon_92} W. von Ammon, Nucl. Instr. and Meth. B 63 (1992)
95.

\bibitem{Shockley_38} W.Shockley. Journ. Appl. Phys., vol.9, 635,
(1938)

\bibitem{Ramo_39} S. Ramo, Proc. I.R.E. 27 (1939) 584.

\bibitem{Hunsuk_91}Hunsuk Kim, et al., Solid-State-Electronics, vol.34,
no.11, 1251, (1991).

\end{thebibliography}
\end{document}